\documentclass[secnumarabic,amssymb,nobibnotes,aps,prd,showpacs]{revtex4}%
\usepackage{graphicx}
\usepackage{dcolumn}
\usepackage{bm}
\usepackage{amsmath}%
\usepackage{amsfonts}%
\usepackage{amssymb}

\begin{document}

\title{Signals at ground level of relativistic solar particles associated to the
``All Saints''  filament eruption on 2014
}

\author{C. R. A. Augusto, C. E. Navia, M. N. de Oliveira and H. Shigueoka}
\affiliation{Instituto de F\'{\i}sica, Universidade Federal Fluminense, 24210-346,
Niter\'{o}i, RJ, Brazil} 

\author{Andr\'{e} Nepomuceno}
\affiliation{Departamento de Ciências da Natureza, IHS, Universidade Federal Fluminense}

\author{A. C. Fauth}
\affiliation{Instituto de F\'{i}sica Gleb Wathagin, Universidade Estadual de Campinas, Campinas, SP Brazil}


\date{\today}

\begin{abstract}
 
Far away from any sunspot, a bright flare erupted on November 1st, 2014, with onset at 4:44 UT and a duration of around three hours, causing a C2.7-class flare. The blast was associated with the sudden disappearance of a large dark solar filament. The rest of the filament flew out into space, forming the core of a massive CME. Despite the location of the explosion over the sun's southeastern region (near the eastern edge of the sun) not be geoeffective, a radiation storm, that is, solar energetic particles (SEP) started to reach the Earth around 14:00 UT, reaching the condition of an S1 (minor) radiation storm level on Nov. 2th. In coincidence with onset of the S1 radiation storm (SEP above 5 MeV), the Tupi telescopes located at  $22^090'$S; $43^020'$W, within the South Atlantic Anomaly (SAA) detected a muon enhancement caused by relativistic protons from this solar blast.
In addition an increase in the particle intensity was found also at
South Pole neutron monitor. This means that there was a transverse propagation to the interplanetary magnetic field of energetic solar particles. However, we show that perpendicular diffusion alone cannot explain these observations, it is necessary  a combination with further processes as
a very high speed, at least of a fraction the CME shocks, close to the ecliptic plane.

\end{abstract}


\maketitle

\section{Introduction}

The currently solar cycle 24 that started at the beginning of 2008 \cite{svalgaard10},
was an anomalous extended period of minimal solar activity.
This Sun cycle anomaly is the first in the spatial era, i.e., where the Sun is monitored by 
spacecraft detectors,  but there are registers from 107 years ago of similar pattern that happened during the transition 
between cycles 13 and 14.

The solar activity in the current solar cycle also has not been enough to produce ground level enhancements (GLEs). Years of maximum solar activity have produced few particle excess at ground level linked with radiation storms, and in most cases with a intensity variation below the 4\%, even in NM located at polar regions.

It is well established that solar flares and its associated CMEs are the origin of high energy radiation, such as gamma and X-rays, energetic particles and interplanetary (IP) shocks that travels through our solar system and interacts with everything it comes across, including the Earth's magnetic field.

Space weather is monitored by NASA and other agencies to alert us when solar activity that starts 92 million miles away might affect life on the ground. However, not all solar events observed in satellites produce any effect at ground level, it is also necessary monitor those solar events that give some sign on the ground, probably these events have more influence in the weather conditions. Thus measurement on satellites and on the ground level are complementary. This is the aim of the Tupi muon telescopes, the study of the effect at  ground level of the space weather.

On the other hand, solar filaments are large regions of very dense, cool gas, held in place by magnetic fields
\cite{guo10}. They usually appear long and thin above the chromosphere, and as they are cooler than the surroundings, they appear dark. But solar filaments on the "edge" of the Sun, looks brighter than the dark outer space behind them.

From end to end, a solar filament stretches more than a million km or about three times the distance between Earth and the Moon. Sometimes the filament becomes unstable, collapse, and create an enormous wall of plasma rises towering over the sun's surface, triggering a Hyder flare. The plasma ejected out into space forms the core of a massive CME.

This phenomenon happened on November 1st, 2014 when a Hyder flare began
at 4:40 UT. A brief report on this event is available at STCE Newsletter ($http://www.stce.be/newsletter/pdf/2014/STCEnews20141107.pdf$). In general, the Hyder flares are longer than the flares from magnetic activity of sunspots. In the present case, the Hyder flare had a duration of three hours, and its maximum intensity reached the condition of a C2.7-class flare. In addition, due to  location of the filament at southeastern of the solar disc, the CME ejected was not  Earth directed.

 In section 2, a brief report on location and setup of the Tupi telescopes is presented. We investigate the hypothesis of a local geomagnetic effect at central region of the SAA since this region has an anomalous weak geomagnetic field strength, the lowest on the world.
We show that in the SAA region cosmic ray fluxes at lower energies are even higher than world averages at comparable altitudes.  In order to see why the SAA region is a good place to make observations of solar transient events, an analysis on the population of energetic particles on the SAA is presented.

Section 3 is devoted to show the observations from space and at ground level of this peculiar Hyder flare event. We found a good correlation between the radiation storm, particularly SEP above 5 MeV, and ground level observation. Both have approximately the same onset.

In order to see how solar energetic particles (SEP), from a solar blast close to the eastern edge of sun, i.e. not geo-effective, it has arrived to the earth, some processes, such as the transverse propagation, including drift processes and open magnetic field tubes are considered in section 4. However,  none of these processes alone take into account the observations. Perpendicular diffusion, including drift processes may be the explanation for these
observations,  however in combination with further essential  processes, such as the existence of CME shocks with high velocities near the ecliptic plane.

The above result comes from two cross checks, made in section xix, the first between the catalogue from NOAA of Solar Proton Events (SPE) (http://www.swpc.noaa.gov/ftpdir/indices/SPE.txt) listing from 1976 to the present and the  catalogue from OULU of Ground level events (GLE)(http://cosmicrays.oulu.fi/GLE.html) listing from 1966 to the present. The aim is to obtain a correlations between the location of the solar active region (triggering radiation storms) and GLEs.

The second cross check was betwen the OULU GLE catalogue and the CACTus COR2 CME list, listing from April 1997. This last cross check is extremely important for our analysis, because it provides  inquire under what circumstances a CME can trigger a GLE.

In section 5 a spectral analysis is carried out through a hybrid method. From this analysis is possible to see that a very hard proton energy spectrum happened (in the GeV energy region) at time of the  muon peak. A comparison with the integral GOES13 proton flux is also performed. Finally in section 6 we present our conclusions.

\section{Location and experimental setup of the Tupi muon telescopes}

\subsection{Experimental setup}

\begin{figure}
\hspace*{-0.0cm}
\center
\vspace*{-3.0cm}
\includegraphics[width=5.5in]{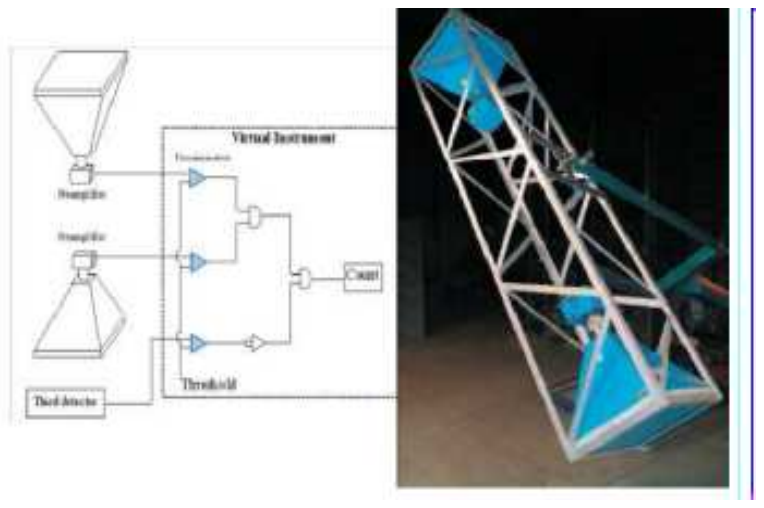} 
\vspace*{-10.0cm}
\caption
{Left:General layout of a Tupi telescope including
the logic in the data acquisition system. Right: Photograph of
the inclines 45 degree West Tupi telescope.}
\label{fig1}
\end{figure}

Since 2013 August, the Tupi experiment has been operating
an extended array of five muon telescopes. Some details are available at
\cite{augusto15}. The first one has a vertical orientation. The other four
have orientations to the north, south, east, and west, and each
telescope is inclined $45^0$ relative to the vertical. 
Each telescope was constructed on the basis of two detectors
(plastic scintillators (50 cm x 50 cm x 3 cm) separated by a
distance of 3 m; one of them is shown in Figure~\ref{fig1}.
Each telescope counts the number of coincident signals in
the upper and lower detector. The output raw data consists of a
coincidences counting rate of 1 Hz versus universal time (UT).
The Tupi telescopes are placed inside a building under two
flagstones of concrete (150 g cm$^2$). The flagstones increase the
detection muon energy threshold up to the $\sim 0.1-0.2$ GeV
required to penetrate the two flagstones. Each Tupi telescope
has an effective FOV of $\sim 0.37$ sr. To the vertical telescope, this
corresponds to an aperture (zenith angle) of $20^0$ from the
vertical.

All steps from signal discrimination to the coincidence
and anticoincidence are made via software using the
virtual instrument technique. The application programs were
written using the LAB-VIEW tools. The Tupi experiment has a
fully independent power supply, with an autonomy of up to
6 hr to safeguard against local power failures. As a result, the
data acquisition is basically carried out with a duty cycle of
95\%. The Tupi experiment is in the process of constant
expansion and upgrade. Work is underway to setup new
telescope sites in Campinas (Brazil) and La Paz (Bolivia).
The last Tupi results on solar transient events is available at \cite{augusto12}.

\begin{figure}
\hspace*{-0.0cm}
\center
\vspace*{-1.0cm}
\includegraphics[width=3.5in]{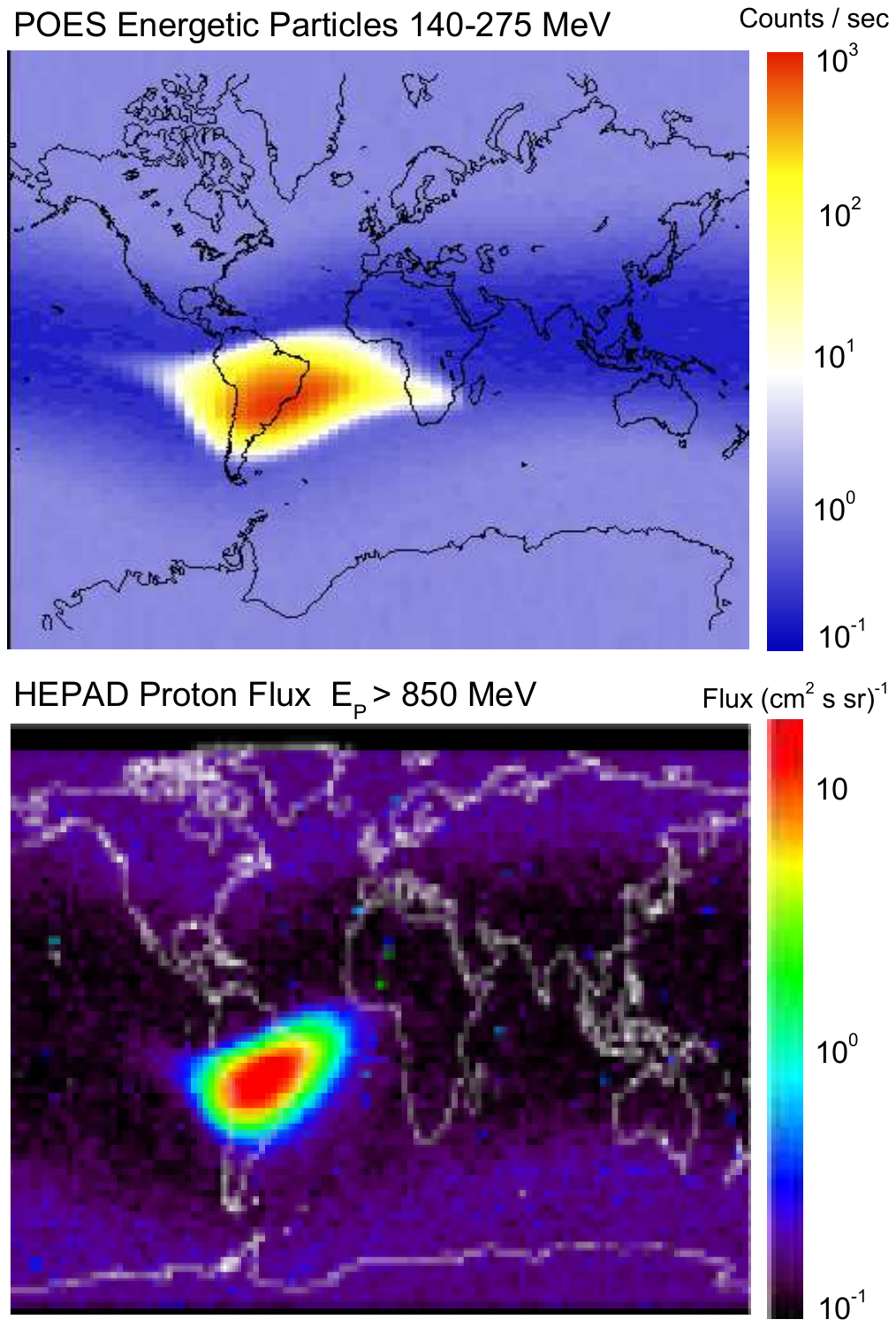} 
\vspace*{-0.0cm}
\caption
{Geographical distribution (latitude vs longitude) of particle counting rate ($140-275$ MeV) measured by the POES satellite (top panel), and the 
proton flux ($E>850MeV$) measured by the HEPAD ICARE detector\cite{boatella10}(bottom panel). 
}
\label{fig2}
\end{figure}

\subsection{Energetic particles in the SAA}

The intensity and the spectrum of SEP observed by ground-based detectors depend on the locations of these detectors on Earth, 
as well as on the time of the occurrence of solar flares. 
In most cases and due to their low geoamagnetic rigidity cutoff, ground-based detectors positioned in polar regions are able to register SEP from solar flares.  Thus, the polar regions are the most favourable locations to detect solar transient events. 
Ground-based detectors positioned near the geomagnetic equator in most cases are usually unable to register SEP from solar flares due to high geomagnetic cutoff rigidity. 

However, there is one more region on Earth, where the Earth magnetic field is weakened considerably (in at least two times at same latitude) \cite{barton97}. 
This is the South Atlantic Anomaly (SAA) at mid-South latitudes, off the coast of Brazil.

 In 1958 Van Allen with Explorer I and II observed that particle detectors stopped to work at altitudes exceeding 2000 km \cite{baker04}. Particle detectors saturated because they could not bear the increasing flux. Van Allen had discovered two radiation belts around Earth containing trapped particles: the inner belt mainly composed by protons and the outer belt rich in electrons.
 
The SAA is the region where the inner Van Allen belt reach very low altitudes, forming a kind of hole in the magnetosphere. In addition,
the condition of stable trapped particles  must be satisfy the Alfven criterion
\begin{equation}
\frac{\rho_L}{\rho_m} << 1,
\end{equation}
where $\rho_L$ is the Larmor radius and $\rho_m$ is the curvature radius of the magnetic field line. Cosmic rays and solar particles with energies below 200 MeV satisfy the above criterion and they can be trapped in the Van Allen belts, for instance, protons can be trapped in the inner belt, following a gyration motion about magnetic field lines, bouncing between the two hemispheres. 
As the inner belt reaches very low altitude at SAA, the result is a high flux (in up to thousand times) of charged particles in this region, such as measured by several spacecrafts. The geographical distribution of the POES \cite{edwards00} omnidirectional of energetic particles counting rate with energies 140-245 MeV is shown in Figure~\ref{fig2} (top panel). The place where the counting rate present a maximum value is the SAA. 

On the other hand, the geographical distribution of proton flux measured by the HEPAD ICARE instrument on-board the Argentinean satellite SAC-C is shown in Figure~\ref{fig2} (bottom panel)
\cite{boatella10}. An excess (at least 10 times) of protons with energies above 850 MeV can be seen in the SAA central region in comparison with the region outside of the SAA. 
These high energy protons are hard to be considered as Van Allen trapped protons, 
above these energies, they no longer satisfy the Alfven criterion of trapped particles. In addition, the SAA models such as AP8 and several measurements of the trapped protons show that their energies do not exceed 300 MeV \citep{boatella10}. 

From this analysis we can conclude that there is an enhancement in the counting rate at detectors within the SAA region, due to the incoming primary cosmic ray and solar particle flux. This characteristic, together with the detection of secondary particles with a low energy threshold, can be crucial to the observation
of transient solar events including those of small scale.

\section{Observations}

\begin{figure}
\vspace*{-0.0cm}
\hspace*{0.0cm}
\centering
\includegraphics[width=14.0cm, angle=90]{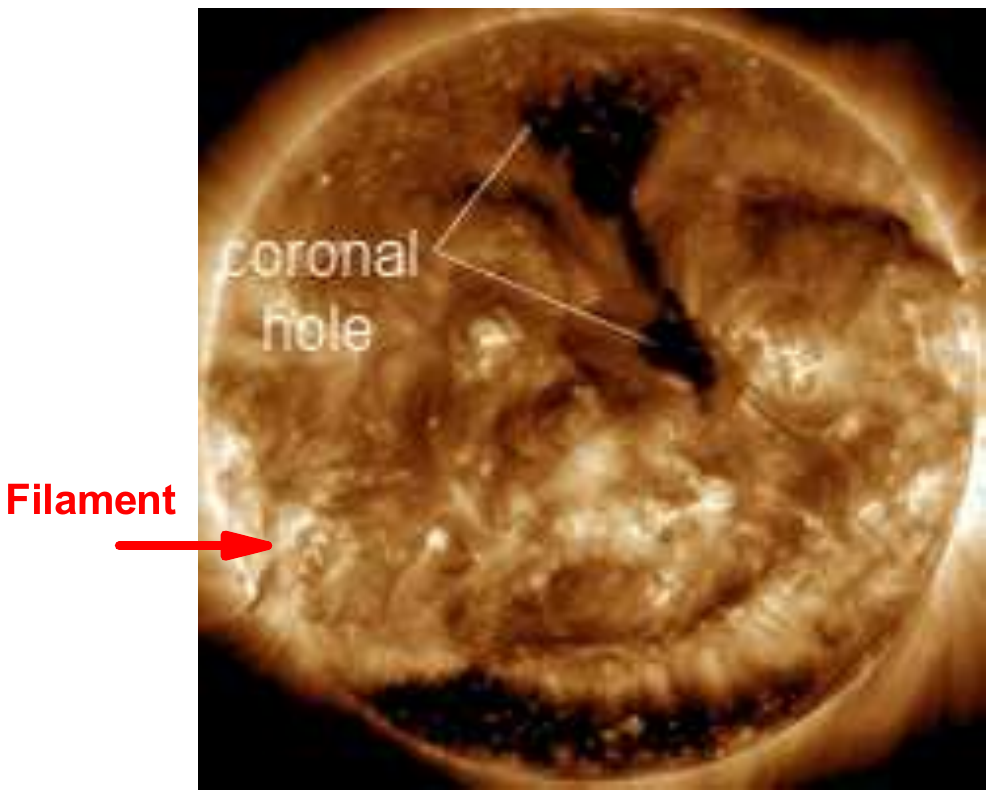}
\vspace*{-4.0cm}
\caption{SDO/AIA observations of coronal holes on October 31, 2014
The arrow at South East of the solar disc indicates the filament
that gave origin to the solar blast, a C3.0-class Hyder flare, on November
1, 2014.}
\label{fig3}
\end{figure}

During the last days of October, a filament at the southeastern region of the solar disk was
observed. Nothing special, if compared to the huge coronal dark holes on the north side of the disk, such as shown in the SDO/AIA observations and reproduced in Figure~\ref{fig3}.

\begin{figure}
\vspace*{-2.0cm}
\hspace*{0.0cm}
\centering
\includegraphics[width=12.0cm]{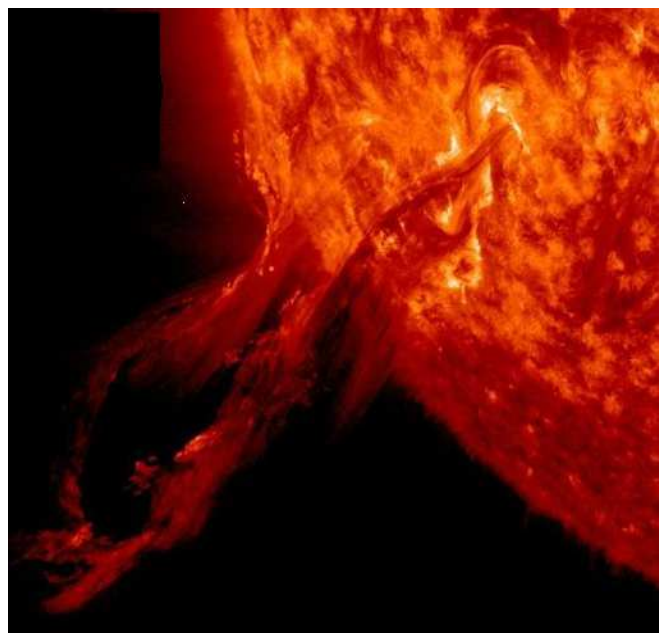}
\vspace*{-8.0cm}
\caption{Image from Solar and Heliospheric Observatory, showing a CME ejected away from the sun, the origin was the sudden disappearance of a large dark solar filament, producing a Hyder flare, of C2.7-class, on November, 2014 with onset at 4:44 UT.}
\label{fig4}
\end{figure}  

\begin{figure}
\vspace*{-0.0cm}
\hspace*{0.0cm}
\centering
\includegraphics[width=10.0cm]{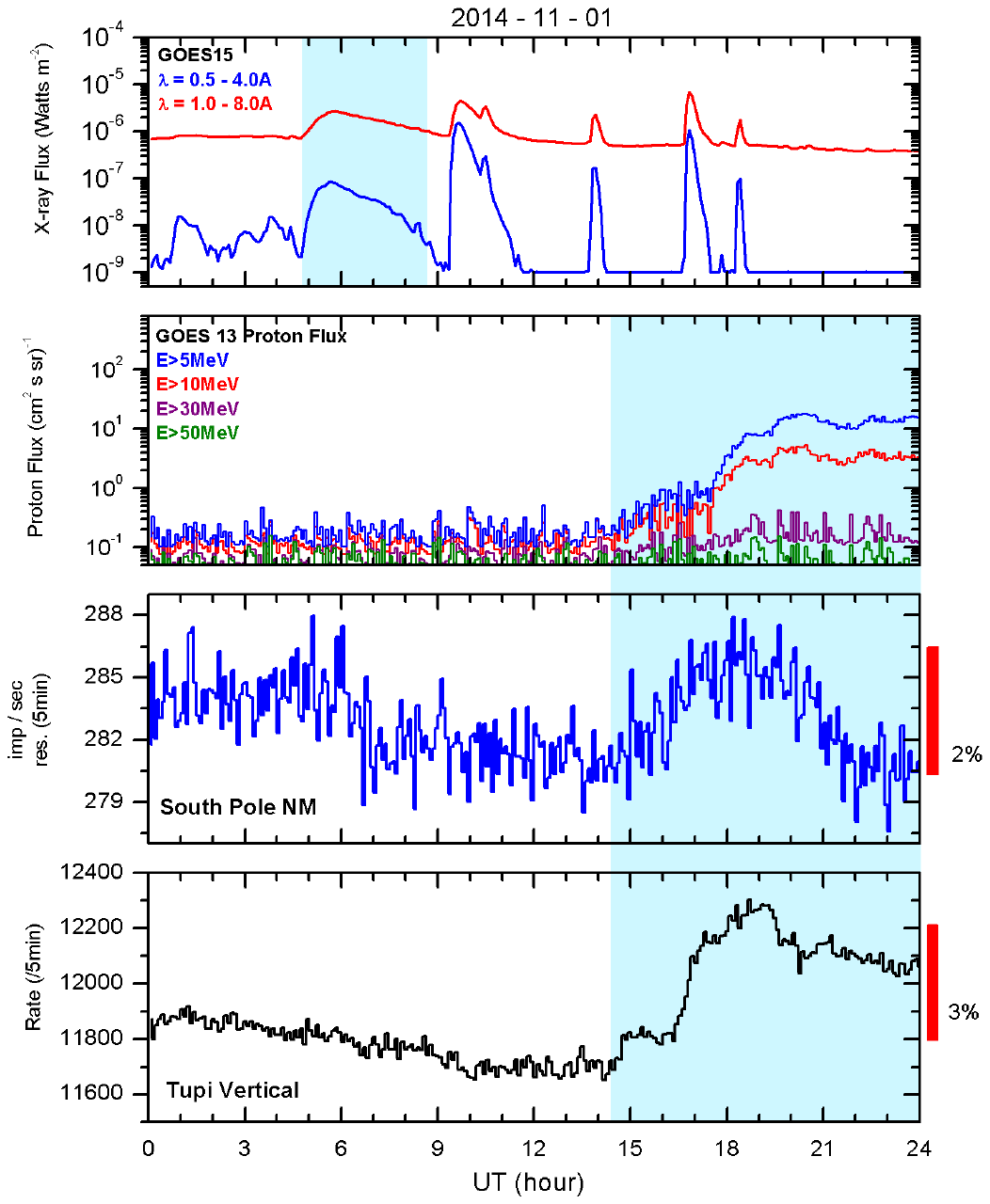}
\vspace*{-0.0cm}
\caption{Time profiles on November 1, 2014, in four different detectors. Top panel: the 
 GOES15 x-ray flux in two wavelength. Second panel: the GOES 13 particle flux in four energy band. Third and bottom panels: the counting rate in the South Pole NM and in the vertical Tupi telescope, respectively.}
\label{fig5}
\end{figure} 

This filament became unstable in the early hours of November 1, 2014, an it burst at 04:40 UT as a flare of C2.7-class, with a duration of around three hours. Flares which happen without sunspots like this are called Hyder Flares. 
The filament flew out into space, forming the core of a massive CME, as shown in the image from Solar and Heliospheric Observatory reproduced in Figure~\ref{fig4}.

According to the X-ray flux detected by GOES 15, on November 1, 2014, there was a flare of class C2.7, with onset at 04:44 UT, peaking at 05:34 UT and end at	07:05 UT, as is shown in Figure~\ref{fig5} (top panel).
As already was commented, there was a CME associated to the blast, but due to the location near the east limb, more precisely at southeastern region, the plasma ejected was not directed towards Earth.

\begin{figure}
\vspace*{-0.0cm}
\hspace*{0.0cm}
\centering
\includegraphics[width=10.0cm]{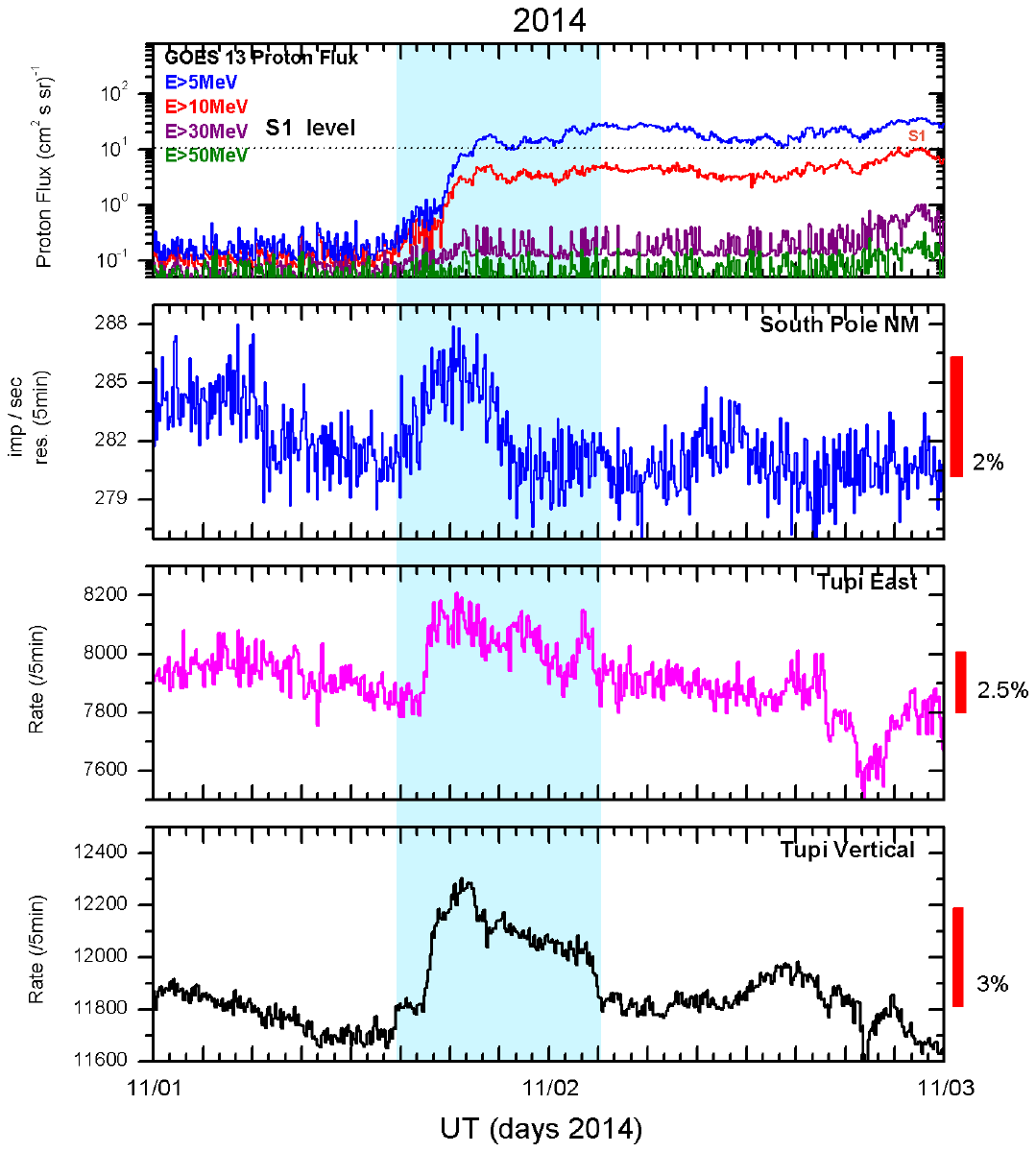}
\vspace*{-0.0cm}
\caption{Time profiles on November 1 and 2, 2014, in three different detectors. Top panel: the 
  GOES 13 particle flux in four energy band. Central and bottom panels: the counting rate in the South Pole NM and in the vertical Tupi telescope, respectively.}
\label{fig6}
\end{figure}

The solar energetic particle (SEP), associated to this blast, began to reach the Earth at 14:10
UT, that is, around 10.5 hours after of beginning of X-ray emission, as is shown in Figure~\ref{fig5} (second panel) where the GOES 13 proton flux is shown. The SEP reached the condition of a radiation storm S1, only in the last hours of November 2th, as shown in the top panel of Figure~\ref{fig6}. These observations suggest that solar particles were accelerates
by CME shocks and the particles were spread over a broad range of longitudes.

In addition, the large delay between the X-ray and the solar energetic particles observed at 1 AU, and the  location of the blast at the southeastern of the solar disk, without a direct magnetic connection imply that the propagation of the SEP was transverse to the interplanetary magnetic field. This hypothesis is described with some detail in section xix.

There is also evidence that the CME's shocks have accelerated particles (protons) up to relativistic energies, that is, above GeV energies, producing particle showers in the Earth atmosphere. One of this evidence is the observation of an increase on the counting rate of ground level detectors, such as the South Pole Neutron Monitor, as well as, in the muon counting rate in the Tupi telescopes. The onset of these counting rate increasing at ground level is in coincidence with the onset of the SEP (above 5 MeV) observed by GOES 13. Figure~\ref{fig5} and ~\ref{fig6}, summarize the situation.

The low confidence level of the counting rate, less than 3\%, at NMs stations, could explain why there was no ground level enhancement (GLE) alarm, such as the  Bartol Neutron Monitors GLE alarm \footnote{$http://www.bartol.udel.edu/~takao/neutronm/glealarm/index.html$}.

\section{Transverse interplanetary diffusion}

How, solar energetic particles (SEP), from an filament close to the eastern edge of sun, spread over a wide range of longitudes and transversely to the interplanetary solar magnetic field?
Many possible scenarios can be suggested. 
Some of them are the following:

a) The SEP observations in a broad range of longitudes could 
imply the existence of coronal shocks extending at least ~300 degrees \cite{cliver95}
and IP shocks up to ~180 degree at 1 AU  \cite{cliver96}. 
However, in the case of the Hyder flare on Nov. 1st, the angular width of the CME was only 130 degrees, and covering only the eastern region, as shown in Figure~\ref{fig8} (left panel). It represents the Lasco C2 image of the CME 0001 on November 1, 2014 at 06:00:06 UT. 

b) Open magnetic field tubes, generated close to the
 source region of the blast, the particles could get away 
from the coronal region of sun following these open magnetic tubes \cite{klein08}. 
This scenario (coronal transport) is equivalent to consider the existence of an almost, direct magnetic connection between the source and the regions of observations.
 However, in the present case, the large delay ($\sim 9$ hours) between the onset of the solar eruption and the onset of the detection of SEP at earth constrain 
the argument of an almost direct magnetic connection.

c) In 1977 Jokipii et al \cite{jopikii77} had already shown that 
the drift velocity of the propagation of galactic cosmic ray with rigidities 
greater than 0.3 GV in the interplanetary medium is around the local solar wind 
velocity. Thus, the drift velocity on the transport equation cannot be neglected. 
This mechanism has been applied to the solar particle propagation with energies 
up to 100 MeV by Dalla et al. cite{dalla13}. This scenario could explain at least 
in part, the detection of particles with relativistic energies from a solar blast, 
in a broad longitude range.

d)  Transverse interplanetary diffusion. This scenario require a large ratio between transverse to parallel diffusion coefficients $\kappa_{\perp} / \kappa_{\parallel} \sim 0.5$. Indeed, there are in the literature some predictions claiming that higher-energy particles are less tied 
to field lines than lower-energy particles. A higher transverse diffusion coefficient means they travel more easily from one field line to the next \cite{giacalone99}. In addition, a 3D perpendicular diffusion calculation in the inner heliosphere has been performed by Drogue et al. \cite{drogue10}, including mean free path scales with the gyro radius of the solar particles (4 MeV), and pitch-angle-dependent diffusion perpendicular to the magnetic field.

Indeed, the transverse diffusion including drift processes and the pitch-angle-dependent diffusion are promising scenarios in the high energy region. However, still remain some experimental observation that does not fit within these scenarios. 
For instance, we have analysed several solar blasts triggering CME at the same region as the Hyder flare on Nov. 1st, that is, close to the eastern solar limbo, none of these events have triggered SEP on earth. This means that the diffusion
processes (above commented) may be the explanation for these
observations, if it is combined with further processes.

In order to verify what are these further processes, two cross checks were performed, the first is between the Oulu NM data for GLEs and the the NOAA SPE list. The second is
between the Oulu NM data for GLEs and the CACTus COR2 CME list.

\begin{figure}
\vspace*{-15.0cm}
\hspace*{0.0cm}
\centering
\includegraphics[width=19.0cm]{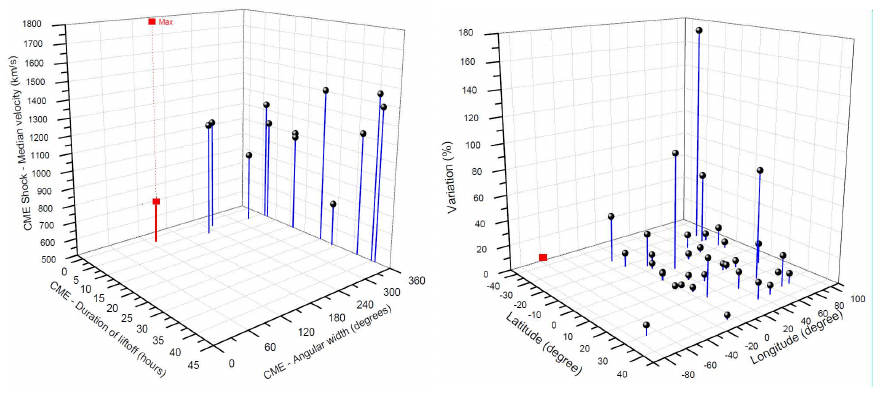}
\vspace*{-4.0cm}
\caption{ Left panel: Lasco C2 image of the CME 0001 on November 1, 2014 at 06:00:06 UT.
The angular width of the CME was 130 degrees with a duration of liftoff of 3 hours.
Right panel:  Velocity distribution of the CME associated shocks, 
as a function of the principal angle counterclockwise from North (degrees) }
\label{fig7}
\end{figure}

\begin{figure}
\vspace*{-3.0cm}
\hspace*{0.0cm}
\centering
\includegraphics[width=14.0cm]{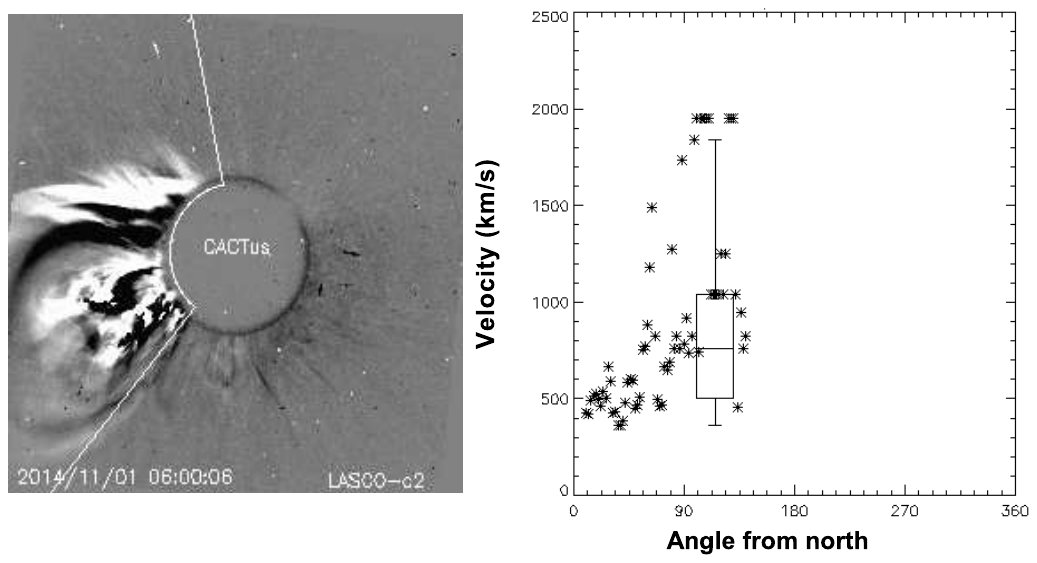}
\vspace*{-8.0cm}
\caption{ Left panel: Lasco C2 image of the CME 0001 on November 1, 2014 at 06:00:06 UT.
The angular width of the CME was 130 degrees with a duration of liftoff of 3 hours.
Right panel:  Velocity distribution of the CME associated shocks, 
as a function of the principal angle counterclockwise from North (degrees) }
\label{fig8}
\end{figure}

The first correlation is shown in Figure~\ref{fig7}(right panel). This correlation means that the solar active regions are located in a restricted band of latitudes, between -45 to 45 degrees. However, they occur in all band of longitude, but in most cases only those explosions from active regions, in the western hemisphere can trigger GLEs. This result is expected due to the sun-earth magnetic connection, due to the Parker spiral behaviour of the magnetic field, on the ecliptic region.
The exception was the GLE29 on September 24th, 1977. Its origin comes from a solar active region, whose coordinates were 30N,70E and with a peak increase of 7\% (resolution of 5 minutes) at OULU NM data. Figure~\ref{fig7} (right panel) summarizes the situation, where the coordinates of the blast, triggering SEP associated with GLEs are plotted as a function of the peak increase of the GLE. For comparison, we have included the coordinates of the Hyder flare on Nov. 1st, represented by the red point.

The second and more important correlation is shown in Figure~\ref{fig7} (left panel).  This correlation can be only reported from April 1997 (begin of the Cactus list) until now,  corresponding to only 17 GLEs, starting with the GLE 55 on November 6, 1977. Thus, only 12 GLEs were correlated with the Cactus CME list \cite{robbrecht09}.
The results show that only full halo CMEs, or at least those with a large angular width, have triggered GLEs. In addition, CMEs triggering GLEs, have
CME shocks with a high median velocity, above 700 km/s. We can see that in most cases this velocity is much higher than 1000 km/s. We can also see that the CME-duration of liftoff is not essential, it is spread in a wide range of value. Again for comparison, we have included the values of the Hyder flare on Nov. 1st, it is represented by the red point.

However, events with high CME shock velocities and CMEs with large angular width are not sufficient to triggering a radiation storms at Earth. It is necessary, high CME shock velocities in the ecliptic plane, as shown in Figure~\ref{fig8}, where the Cactus CME image linked with the Hyder flare on Nov. 1st is show on the left panel and
their CME velocity distribution as a function of the principal angle (in degrees) measured from  counterclockwise from North is shown on the right panel.

\begin{figure}
\vspace*{-10.0cm}
\hspace*{0.0cm}
\centering
\includegraphics[width=14.0cm]{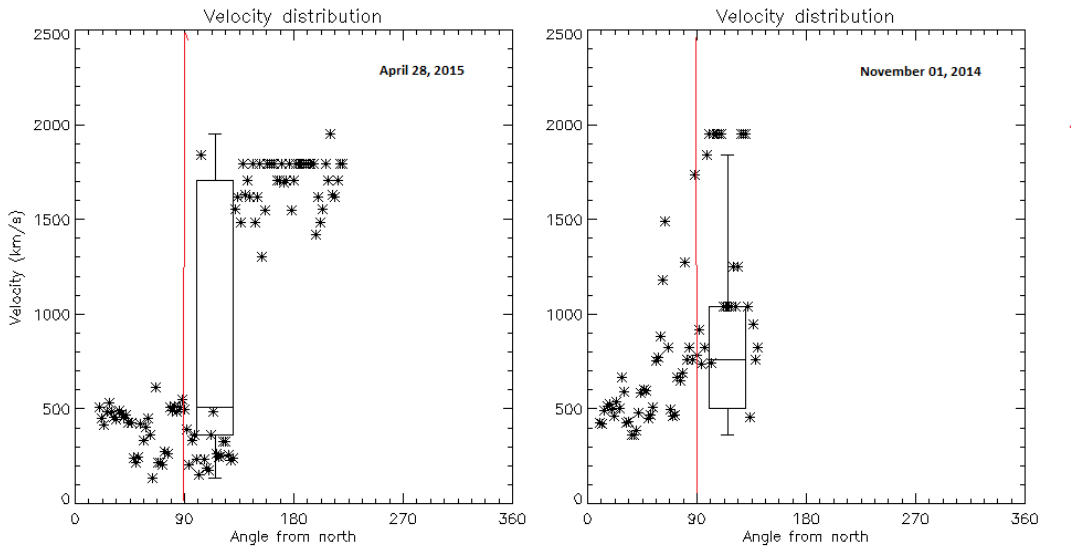}
\vspace*{-1.0cm}
\caption{ Comparison between the CME velocities as function of the principal angle of two similar Hyder flare events, on April 28, 2015 (left panel) and November 1st, 2014 (right panel). Both events have CMEs with high velocities. However, there are CMEs with a principal angle of 90 degrees (ecliptic plane, marked by red line) only at the event on  Nov. 1st. }
\label{fig9}
\end{figure}

We have arrived to this conclusion analysing several events located like the Hyder flare on Nov 1st. An example is shown in Figure~\ref{fig9}, where we compare the CME linked to the Hyder flare on April 28, 2015 with the CME linked with the Hyder flare on November 1, 2014. In both cases the filaments were located at the East solar limb. In addition, in both cases there are high CME shock velocities. However, basically there are CME shock injections with a principal angle (90 degrees), that is, in the ecliptic plane, only in the case of the event on Nov. 1st. We believe that there is transverse propagation only at the ecliptic plane because particles injected out of this plane are directed  at high heliographic latitudes by the  open solar magnetic field lines, the so called  Fisk heliospheric magnetic field.

\subsection{Spectral analysis}

\begin{figure}
\vspace*{-1.0cm}
\hspace*{-1.0cm}
\centering
\includegraphics[width=11.0cm]{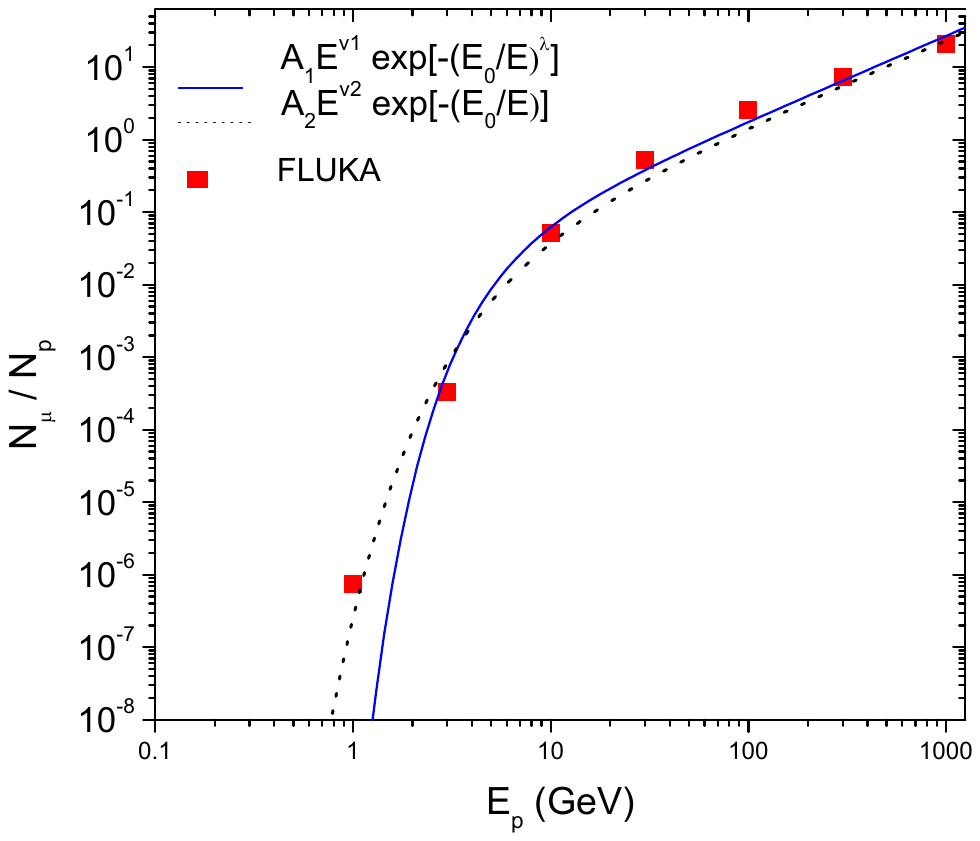}
\vspace*{-6.0cm}
\caption{Yield function, as the number of muons at the sea level per proton (vertical incidence), as a function of incident proton energy, from FLUKA calculations (black squares) \cite{poirier02}. The lines show several fit functions.}
\label{fig10}
\end{figure}

The specific yield function (the number of muons per incident proton on the top of the atmosphere) as a function of proton energy is shown in Figure~\ref{fig10}. It is determined according to the FLUKA \footnote{FLUKA (``FLUktuierende KAskade''), $http://www.fluka.org/fluka.php$}, which is a detailed general purpose tool for calculations of particle transport and interactions with matter. This FLUKA output \cite{poirier02} can be described by the following fit

\begin{equation}
 S(E_P>10GeV)=A_{\mu} E_P^{\nu}exp\left(-(E_{0}/E_P)^{\lambda} )\right),
\end{equation}
where $A_\mu = (7.8 \pm 1.6) \times 10^{-3}$, $\nu=1.18 \pm 0.24$, $E_0=10.2 \pm 2.1$ GeV, $\lambda=1.48 \pm 0.30$. A similar function can be obtained to $E_P<10GeV$; in this case, the parameters are: $A_\mu = (6.8 \pm 1.4) \times 10^{-3}$, $\nu=1.18 \pm 0.24$, $E_0=10.2 \pm 2.05$ GeV, $\lambda=1.00 \pm 0.20$. This fits (solid lines) together with the simulated results, which are shown in Figure~\ref{fig10}.

On the other hand, we assume here that the energy spectrum of solar protons in the GeV energy region, which is in the high energy tail of the SEP spectrum, can be fitted by a single power law function.
\begin{equation}
N_P(E_P)=A_PE_P^{\beta}.
\end{equation} 
There are two unknown quantities in the single parameter power law function: the coefficient $A_P$ and the spectral index $\beta$.

A convolution between the yield function S(E) and the particle spectrum $N_p(E)$ gives the response function, which is the number of muons in the excess signal generated by the SEP during the time period T,
detected by a telescope, as a muon flux $I_{\mu}(>E_{\mu})$.
The convolution can be expressed as

\begin{equation}
I_{\mu}(>E_{\mu})= \int_{E_{min}}^{\infty}S(E_P)A_P E_P^{\beta}dE_P.
\end{equation}

 Furthermore, the integrated time fluence can be obtained as

\begin{equation}
F=T \left[\int_{E_{min}}^{\infty}dE_P A_PE_P^{\beta}E_P\right].
\end{equation}

\begin{figure}
\vspace*{-6.0cm}
\hspace*{-1.0cm}
\centering
\includegraphics[width=11.0cm]{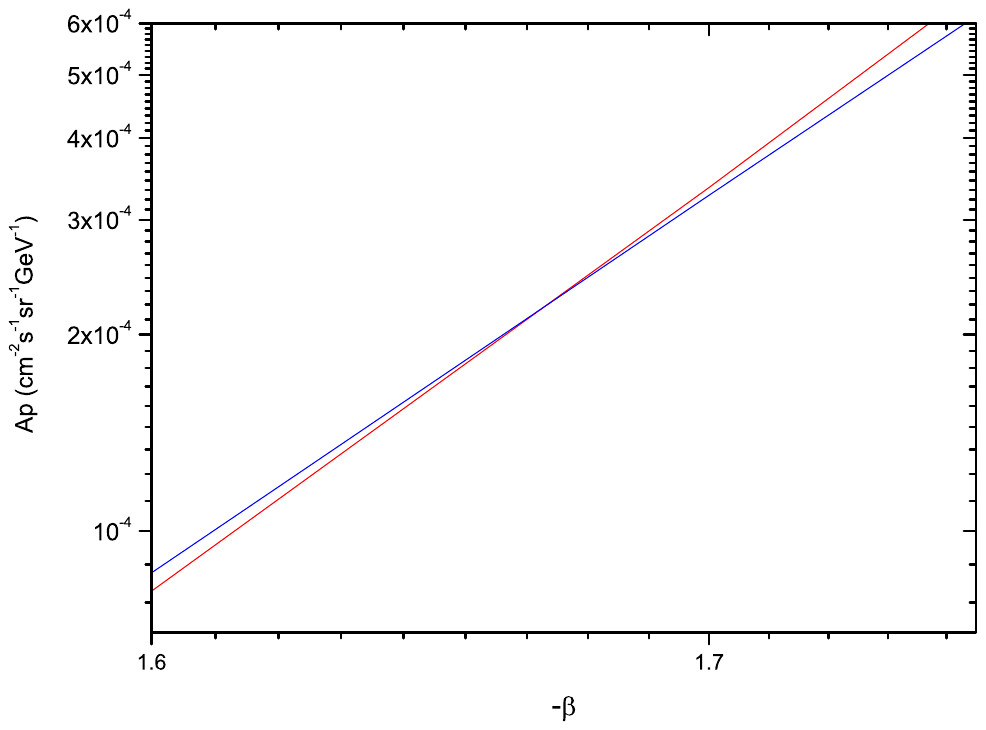}
\vspace*{-0.0cm}
\caption{
Correlation between the coefficient Ap and the spectral index $\beta$. All possible values of Ap and $\beta$ compatible with the observed muon flux (dot line)
and the integrated fluence F (solid line) are obtained on the basis of Monte Carlo simulations and analytical calculations. These quantities are defined by
Eq. 4 and Eq. 5.
}
\label{fig11}
\end{figure}

The terms on the left side of Eq.3 and Eq.5 are known. In the present case, they represent the observed flux and  fluence of the muons at peak. Thus, we can consider all possible values of 
$\beta$ and $A_P$ compatible with the observed muon flux excess value 
$I_{\mu}=(1.25 \pm 0.25)\times 10^{-2}(m^2sr\;s)^{-1}$, and considering that the muon's energy threshold is 0.1 GeV,  one can estimate the observed integrated fluence as at least $F=(0.14 \pm 0.03) GeV/cm^2$. Figure~\ref{fig11} shows that one can obtain the best estimate for the spectral index using the intersection of two lines defined by Eq. 3 and Eq. 5.

\begin{figure}
\vspace*{-6.0cm}
\hspace*{-1.0cm}
\centering
\includegraphics[width=11.0cm]{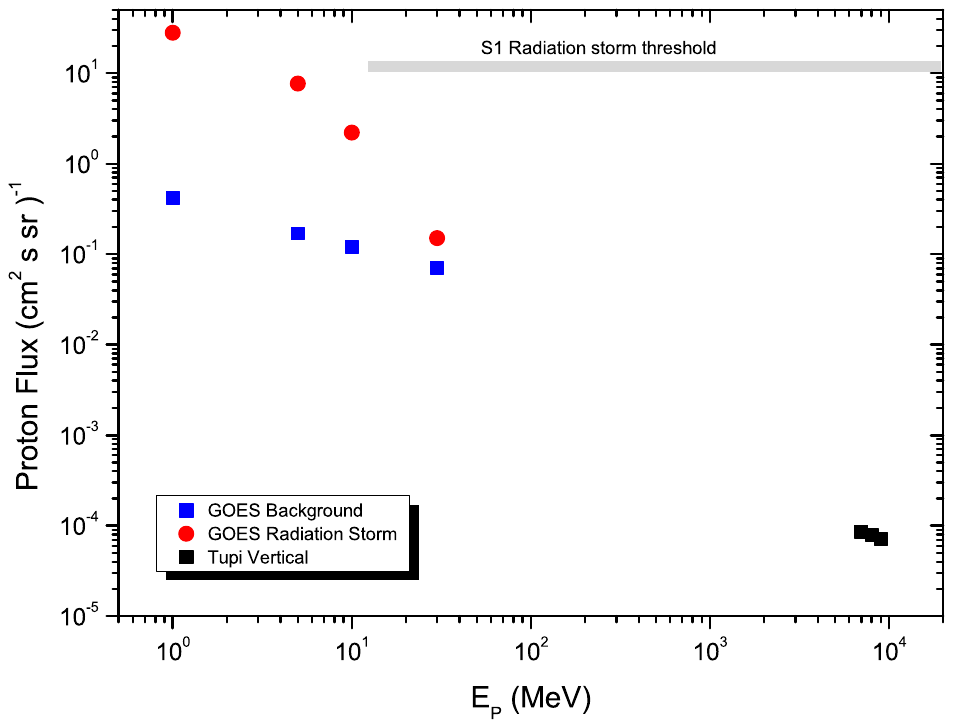}
\vspace*{-0.0cm}
\caption{
Integral proton flux: the blue triangles and the red circles represent the GOES 13 data, and corresponds  to the proton flux background and when the radiation storm reaches the level S1 on Apr 18, 2014, respectively. The black squares represent the proton flux obtained from the muon excess on Tupi telescope, observed in coincidence with the radiation storm.
}
\label{fig12}
\end{figure}

From this analysis, we can find out that the best estimate for the spectral index is compatible with $\beta=-1.67\pm 0.33$ and $A_p=(2.21\pm 0.44)\times 10^{-4} (cm^2 s sr GeV)^{-1}$, both at 68\% confidence level. In order to make a comparison with satellite data, the integral proton flux has been obtained as
\begin{equation}
I_P(>E_P)=\int_{E_p}^{\infty} A_P E_P^{\beta}dE_P.
\end{equation}
The results of the integral proton flux obtained from the Tupi muon excess (at peak), observed in coincidence with the radiation storm, is shown by black squares in Figure~\ref{fig12}. The red circles and the blue triangles represent the GOES 13 data at peak of the Tupi muon flux.

\section{Conclusions}

We have reported a description and an analysis of 
effects at ground level of a blast associated with the sudden disappearance of a large dark solar filament on November 1, 2014.
The explosion initially seen as a Hyder C2.7-class flare, had an onset at 4:44 UT, and a radiation storm, that is, solar energetic particles above 5 MeV started to reach the Earth around 14:00 UT, reaching the condition of an S1 (minor) radiation storm level on Nov. 2th.

In addition, a particle enhancement has been observed at Tupi telescope as a muon excess. The onset of the muon excess observed with a confidence level of 4\% is in temporal coincidence with the onset of the radiation storm observed by GOES13. A similar result is also observed in the counting rate of the South Pole NM.

This is a rare event observed at ground level by two reasons. The first is that the blast is not correlated with sunspots, the origin was a filament. The second is the location of the filament, near the eastern edge of the sun, that is, without direct magnetic connection with the Earth. From a systematic analysis we conclude that indeed the transverse diffusion, including drift processes and the pitch-angle-dependent diffusion are promising scenarios to explain the observations, however, it is necessary to include an important condition, the existence of high speed CME shocks close to the ecliptic plane. Without this condition, the high speed CME shocks injects particles only in regions with an high helio-latitude.

We believe that others favourable conditions, such as the low energy threshold of the muons detected (100 MeV); the location of Tupi telescopes, close to the central region of the SAA, a region where the geomagnetic field is the lowest on the world and specially a very hard proton spectrum, at the time in that the muon excess reaches the peak, inferred through an hybrid method and contributed to observe this event associated with relativistic particles emitted by a CME linked with the Hyder flare on Nov. 1st, 2014.

\acknowledgments

This analysis has made use of NOAA's Space Weather Prediction Center (SWPC), as well as, the ESA Neutron Monitor Service. We would like to thanks to Oulu NM data for the Ground Level Enhancements (GLE) and the CACTus LASCO CME catalog for the data.
This work is supported by the National Council for Research (CNPq) of
Brazil, under Grant 306605/2009-0 and Fundacao de Amparo a Pesquisa do Estado do Rio de Janeiro (FAPERJ), under Grant 08458.009577/2011-81 and E-26/101.649/2011.


\newpage

\begin{references}

\bibitem{svalgaard10}L. Svalgaard and E. W. Cliver,  J. Geophys. Res., 115, A09111 (2010). 

\bibitem{guo10}T. Guo et al., \apj, 714, 343 (2010).

\bibitem{augusto15}C. R. A. Augusto et al., \apj, 805, 69 (2015).

\bibitem{augusto12}C. R. A. Augusto et al., \apj, 759, 143 (2012). 

\bibitem{barton97}C. E. Barton, J. Geomagn. Geoelectr., 4942002, 123 (1997).

\bibitem{baker04}D. N. Baker, S. G. Kaneka, X. Li, S. P. Monk, J. Goldstein and J. L. Burch, NATURE, 432, 878 (2004).

\bibitem{edwards00}P. G. Edwards and D. Pawlak, ESA bulletin 102 may 2000. 

\bibitem{boatella10} C. Boatella, C. et al.\ 2010, IEEE Transactions on Nuclear Science, 57, 2000, (2010). 



\bibitem{cliver95}E. W. Cliver, et al., Proceedings of the 24th International Cosmic Ray
Conference, vol. 4, pp. 257-260, 1995.

\bibitem{cliver96}E. W. Cliver and H. V. Cane,  J. Geophys. Res. 101, 15533 (1996).

\bibitem{klein08}K. L. Klein, S. Krucker, G. Lointier, A. Kerdraon,  Astron. Astrophys. 486, 589 (2008).

\bibitem{jopikii77}J. R. Jopikii, E. H. Levy and W. B. Hubbard, \apj, 213, 861 (1977).

\bibitem{dalla13}S. Dalla, M.S. Marsh, J. Kelly, T. Laitinen, J. Geophys. Res. Space Physics, 118, 1 (2013).

\bibitem{giacalone99}J. Giacalone and J. R. Jokipii, \apj, 520, 204 (1999).

\bibitem{drogue10}W. Droge, Y. Y. Kartavykh, B. Klecker, and G. A. Kovaltsov, \apj, 709, 912 (2010).

\bibitem{robbrecht09}E. Robbrecht et al., \apj, 691, 1222 (2009).

\bibitem{poirier02}J. Poirier and C.  D'Andrea, J. Geophys. Res., 107, 1376 (2002).




\end{references}
\end{document}